\documentclass{article}
\usepackage{graphicx}

\begin{document}

\date{}
\title{Quark Correlations Inside Hadrons And Single Diffraction}
\author{G. Musulmanbekov}


%
\maketitle

\begin{abstract}
Hadron structure is considered in the frame of Strongly Correlated Quark
Model (SCQM). It is shown that quark correlations result in fluctuations of
hadronic matter distributions and single diffractive dissociation processes
in hadronic collisions, hard or soft, are manifestation of these
fluctuations inside colliding hadrons.
\end{abstract}

\section{Introduction}

Elastic and inelastic $pp$ and $\overline{p}p$ $-$interactions
play an important role in understanding of structure of hadrons.
Variety of scattering processes, differentiation them into hard
and soft ones, single and double diffraction and central hadron
production tell us that this structure is rather complicated. What
is the source of complexity? If hard processes are the area of
application of current quark consideration then soft ones are
described in constituent quark approach. Interplay between hard
and soft interactions in high energy experiments strictly relates
to interconnection between small and large size quark
configurations inside a hadron. We think that this interplay is a
manifestation of fluctuations of hadronic matter distributions
inside hadrons and these fluctuations in turn are results of
correlated motion of valence quarks inside hadrons.

The question arises: is it possible to construct the dynamical
system of quarks which can be observed at one instant of time as
constituent (dressed) quarks and at the another one - as current
(bare) quarks? Proposed by author the semiclassical model of
strongly correlated quarks, SCQM, demonstrated how these
configurations could be realized inside hadrons\cite{genis}. In
this paper we give further elaboration of the model which\
possesses the features of both constituent and current quark
models and allows one to extract confining potential and
force(Section 2). In Section 3 we apply this model for description
of diffractive processes and show that explicit manifestation of
quark correlation inside hadrons is a single diffractive
scattering.

\section{Strongly Correlated Quark Model}

Let us imagine the following hypothetical picture: single, colored
quark imbedded into vacuum. Because of vacuum fluctuations one can
observe two competing processes: first, polarization of
$\overline{q}q$ sea by the color field of valence quark (VQ) and
second, the tend of vacuum fluctuations to destroy this
polarization. As a result, one can say about a vacuum pressure on
a single, colored quark. This effect can be interpreted as the
''response'' of the vacuum on the presence of, say, point defect
or dislocation like in solid state physics. What happens if we
place in vicinity of this quark the corresponding antiquark? By
virtue of opposite color signs their polarization fields interfere
in the overlapping region \textbf{destructively}. So the pressure
of vacuum on quark (antiquark) from outside exceeds that one going
from inner space between quark and antiquark. This results in an
\textbf{attractive} force between quark (dislocation) and
antiquark (antidislocation). The density of the remaining part of
polarization field around quark (antiquark) is identified with
hadronic matter distribution. At maximum displacement in
$\overline{q}q-$ system, that corresponds to small overlapping of
polarization fields, hadronic matter distributions have maximum
values. So quark and antiquark located nearby start moving towards
each other. The closer they to one another, the larger destructive
interference effect and the smaller hadronic matter distributions
are around VQs and the larger their kinetic energies.

For such interacting $\overline{q}q-$ system the total Hamiltonian is%

\begin{equation}
H=\frac{m_{\overline{q}}}{(1-\beta^{2})^{1/2}}+\frac{m_{q}}{(1-\beta
^{2})^{1/2}}+V_{\overline{q}q}(2x),
\end{equation}
were $m_{\overline{q}}$, $m_{q}-$ masses of valence antiquark and quark,
$\beta=\beta(x)-$ their velocity depending on displacement $x$ and
$V_{\overline{q}q}-$ quark--antiquark potential energy with separation $2x.$
It can be rewritten as%

\begin{equation}
H=\left[  \frac{m_{\overline{q}}}{(1-\beta^{2})^{1/2}}+U(x)\right]  +\left[
\frac{m_{q}}{(1-\beta^{2})^{1/2}}+U(x)\right]  =H_{\overline{q}}+H_{q},
\end{equation}
were $U(x)=\frac{1}{2}V_{\overline{q}q}(2x)$ is potenial energy of
quark or antiquark. Therefore, keeping in mind that quark and
antiquark are strongly correlated we consider each of them
separately as undergoing oscillatory motion in 1+1 dimension.
Generalization to three--quark system in baryons is performed
according to $SU(3)_{color}$ symmetry: an antiquark is replaced by
two correspondingly colored quarks to get color singlet baryon and
destructive interference takes place between color fields of three
valence quarks. Putting aside the mass and charge differences of
valence quarks we may say that inside baryon three quarks
oscillate along the bisectors of equilateral triangle. Hereinafter
we consider that axis $Z$ is perpendicular to the plane of
oscillation $XY$.

VQ with its polarized surroundings (hadronic matter distribution) form
constituent quark. According to our approach potential energy of valence
quark, $U(x),$ corresponds to the mass $M_{Q}$ of constituent quark:%

\begin{equation}
U(x)=const\int_{-\infty}^{\infty}dz^{\prime}\int_{-\infty}^{\infty}dy^{\prime
}\int_{-x}^{\infty}dx^{\prime}\rho(x,\mathbf{r}^{\prime})\approx M_{Q}(x)
\end{equation}
with%

\begin{equation}
\rho(x,\mathbf{r}^{\prime})=\left|  \varphi(x,\mathbf{r}^{\prime})\right|
^{2}=\left|  \varphi_{Q}(x^{\prime}+x,y^{\prime},z^{\prime})-\varphi_{\overline{Q}%
}(x^{\prime}-x,y^{\prime},z^{\prime})\right|  ^{2}.
\end{equation}
The knowledge of the mechanism and structure of vacuum
polarization around valence quark would give the information about
the confining potential. We cannot say at the moment for sure what
is the microscopical mechanism of interaction of valence quark
with vacuum. It could be instanton induced interactions,
excitation of fractal structure of space--time, etc. So we assume,
as a first approximation, that the polarization field can be taken in gaussian form:%
\begin{equation}
\varphi_{Q}(\mathbf{r})=\varphi_{Q}(x,y,z)=\varphi_{Q}(x_{1},x_{2}%
,x_{3})=\frac{(\det\hat{A})^{1/2}}{(\pi)^{3/2}}\exp\left(  -\mathbf{X}^{T}%
\hat{A}\mathbf{X}\right)  ,
\end{equation}
where exponent is written in quadratic form. The same is$\ $for $\varphi
_{\overline{Q}}(\mathbf{r}).$ We define the mass of constituent quark at
maximum displacement
\begin{equation}
M_{Q(\overline{Q})}(x_{\max})=\frac{1}{3}\left(  \frac{m_{\Delta}+m_{N}}%
{2}\right)  \approx360\ MeV,
\end{equation}
where $m_{\Delta}$ and $m_{N}$ are masses delta--isobar and nucleon
correspondingly. Therefore, the parameters of the model are the masses of
VQs,$\ m_{q(\overline{q})},$ which are chosen to be $5$ $MeV$, maximum
displacement, $x_{\max},$ and parameters of gaussian function, $\sigma
_{x,y,z}.$ As shown below, $x_{\max}$ and $\sigma_{x,y,z}$ are adjusted by
comparison of calculated and experimental values of inelastic cross sections,
$\sigma_{in}(s),$ for $pp$ and $\overline{p}p-$ collisions.
\begin{figure}[tb]
\begin{center}
\includegraphics[width=4.5in]{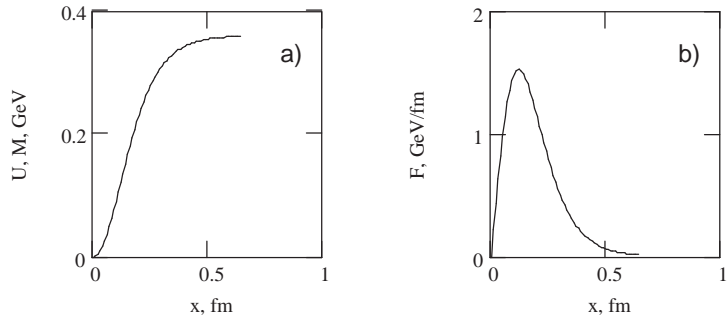}
\caption{a)Potential energy of valence quark and mass of
constituent quark; b)"Confinement" force.}
\end{center}
\end{figure}Using (3)--(5) we can calculate the confining potential $U(x)$ and
force $F(x)=-\frac{dU}{dx}$ . They are shown in Fig. 1. As one can
see the confining potential is essentially nonlinear. The behavior
of potential evidently demonstrates the relationship between
constituent and current quark states inside a hadron. At maximum
displacement quark is nonrelativistic, constituent one (VQ
surrounded by ''polarized sea''), since according to (3) the
confining potential corresponds to the mass of constituent quark.
At the origin of oscillation, $x=0,$ antiquark--quark in mesons
and 3 quarks in baryons, being close to each other, have maximum
kinetic energy and correspondingly minimum potential energy and
mass: they are relativistic, current quarks (bare VQs).
Intermediate region corresponds to increasing (decreasing) of
quark's mass by dressing (undressing) of quarks due to vacuum
polarization. This mechanism agrees with local gauge invariance
principle. Indeed, phase rotation of wave function of single quark
in color space $\psi_c$ on angle $\theta$ depending on
displacement $x$ of the quark in coordinate space
\begin{equation}
  \psi_{c}(x)\rightarrow e^{i\theta(x)}\psi_{c}(x)
\end{equation}
results in it's dressing (undressing) by quark and gluon
condensate that corresponds to the transformation of gauge field
$A_\mu =(\varphi,0,0,0)$
\begin{equation}
  A_\mu(x)\rightarrow A_\mu(x)+\partial_\mu\theta(x).
\end{equation}
Here we dropped color indices and took into account only scalar
component, $\varphi$, of gauge field. Thus gauge transformation
maps internal (isotopic) space of colored quark onto coordinate
space. On the other hand this dynamical picture of VQ dressing
(undressing) corresponds to chiral symmetry breaking
(restoration). The behaviour of field $\varphi$ and hadronic
matter distribution,$\rho$, for quark--antiquark system during
their oscillations is shown in Fig. 2.
\begin{figure}[tb]
\begin{center}
\includegraphics[width=4.7in]{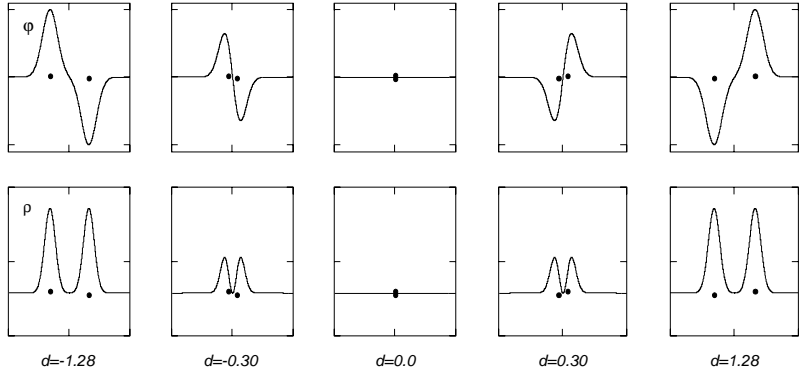}
\end{center}
\caption{Evolution of field $\varphi$, (Eq. (5)), and hadronic
matter distribution $\rho$, (Eq. (4)), in quark--antiquark system
during one--half of the period of oscillations; $d=2x$ -- distance
in {\it fermi} between quark and antiquark depicted as dots.}%
\end{figure}
Due to this mechanism of VQs oscillations nucleon runs over the
states corresponding to the certain terms of the infinite series
of Fock space
\begin{equation}
\mid B\rangle=c_{1}\mid q_{1}q_{2}q_{3}\rangle+c_{2}\mid q_{1}q_{2}%
q_{3}\overline{q}q\rangle+...
\end{equation}

Confining force drastically differs from the one given by string
models. When VQs are close each other it is very weak and fulfills
the ''asymptotic freedom'' behavior of quarks of QCD. At larger
distances between VQs it starts growing rapidly, then reaching
maximum value goes down, asymptotically approaching zero. Thus at
large distances inside hadrons quarks being in a constituent state
are almost free. Hence, it is clear why additive quark model,
where quarks are treated as massive, almost unbound and extended
objects, works well. We must emphasize that interaction between
VQs is not direct but a result of polarization of surrounding
vacuum combined with destructive interference. Attractive force
between VQs in ground state hadrons does not appear as gluon
string but goes from vacuum suppression that predominates from
outside. Therefore, our approach reflects the features of bag
models, as well. The model is in agreement with the experiments
(Fig. 3) for description of VQ
structure function inside a nucleon%

\begin{equation}
F_{2}^{ep}-F_{2}^{en}=\frac{x_{F}}{3}\left[  u_{v}(x_{F})-d_{v}(x_{F})\right]
.
\end{equation}
\begin{figure}[tb]
\begin{center}
\includegraphics[width=3.6495in]{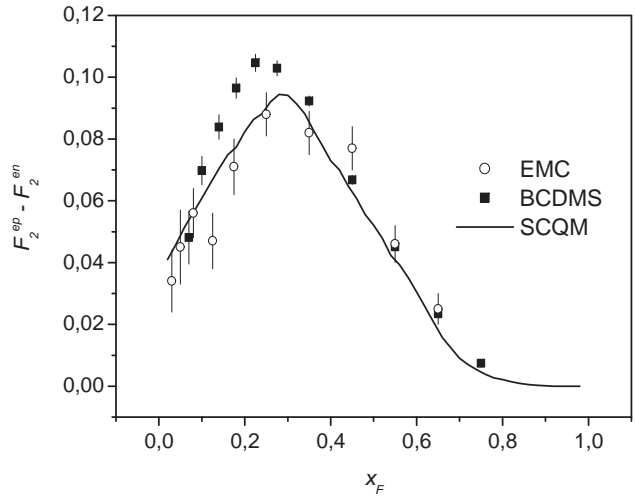}
\end{center}
\caption{Valence quark structure functions in nucleons; data point are from
papers\cite{emc, bcdms}.}%
\end{figure}

Of cause, our description is classical and we must take into
account quantum corrections. It will be the subject of forthcoming
papers. Nevertheless, classical consideration of VQ oscillations
is justified by E. Schrodinger's paper\cite{schrod} where he,
analyzing the motion of wave packet solution of time dependent
Schrodinger equation for harmonic oscillator, demonstrated that
this wave packet moves in exactly the same way as corresponding
classical oscillator. In our model VQ with its surroundings can be
treated as (nonlinear) wave packet. In forthcoming paper we'll
show that these wave packets possess soliton--like features.

\section{Application to Diffractive Processes}

Different configurations of quark contents in colliding hadrons realized at
the instant of collision result in different types of reactions. The
probability of finding any quark configuration inside a hadron is defined by
the probability of VQ's displacement in proper frame of a hadron:%

\begin{equation}
P(x)dx=\frac{Adx}{\sqrt{1-m_{q}^{2}/(E-V)^{2}}}%
\end{equation}
with
\begin{equation}
\int P(x)dx=1.
\end{equation}
Configurations with nonrelativistic constituent quarks ( $x\simeq x_{max}$) in
both colliding hadrons lead to multiparticle production in central and
fragmentation regions. Hard scattering with jet production and large angle
elastic scattering take place when configurations with current VQs ($x\simeq
0$) in both colliding hadrons are realized. Intermediate configurations inside
one (both) of colliding hadron are responsible for single (double) diffraction
processes and semihard scattering. Using impact parameter representation,
namely Inelastic Overlap Function (IOF), we can calculate total, inelastic,
elastic and single diffractive cross sections for $pp$ and $\overline{p}p$
collisions. In impact parameter representation IOF can be specified via the
unitarity equation%

\begin{equation}
2Imf(s,b)=|f(s,b)|^{2}+G_{in}(s,b),
\end{equation}
where $f(s,b)-$elastic scattering amplitude and $G_{in}(s,b)$ is IOF. IOF is
connected with inelastic differential cross sections in impact parameter space:%

\begin{equation}
\frac{1}{\pi}(d\sigma_{in}/db^{2})=G_{in}(s,b).
\end{equation}
\begin{figure}[tb]
\begin{center}
\includegraphics[width=2.2433in]{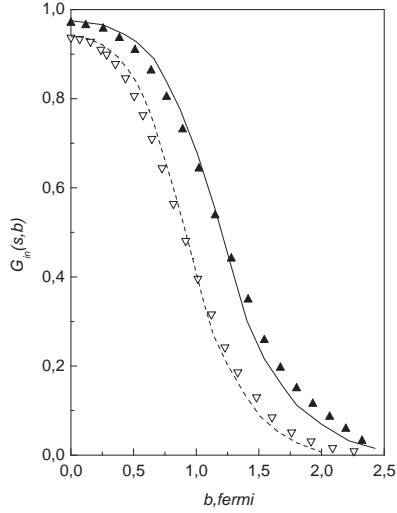}
\end{center}
\caption{Inelastic overlap function for $pp$ and $\overline{p}p-$ collisions
at $\sqrt{s}=53$ and $540$ $GeV$; triangles are Henzi and Valin
approsimation\cite{henzi}.}%
\end{figure}Then inelastic, elastic and total cross sections can be expressed
via IOF as%

\begin{equation}
\sigma_{in}(s)=\int G_{in}(s,\mathbf{b})d^{2}\mathbf{b},
\end{equation}%
\begin{equation}
\sigma_{el}(s)=\int\left[  1-\sqrt{1-G_{in}(s,\mathbf{b})}\right]  ^{2}%
d^{2}\mathbf{b},
\end{equation}%
\begin{equation}
\sigma_{tot}(s)=2\int\left[  1-\sqrt{1-G_{in}(s,\mathbf{b})}\right]
d^{2}\mathbf{b}.
\end{equation}
\begin{figure}[tb]
\begin{center}
\includegraphics[width=4.7in]{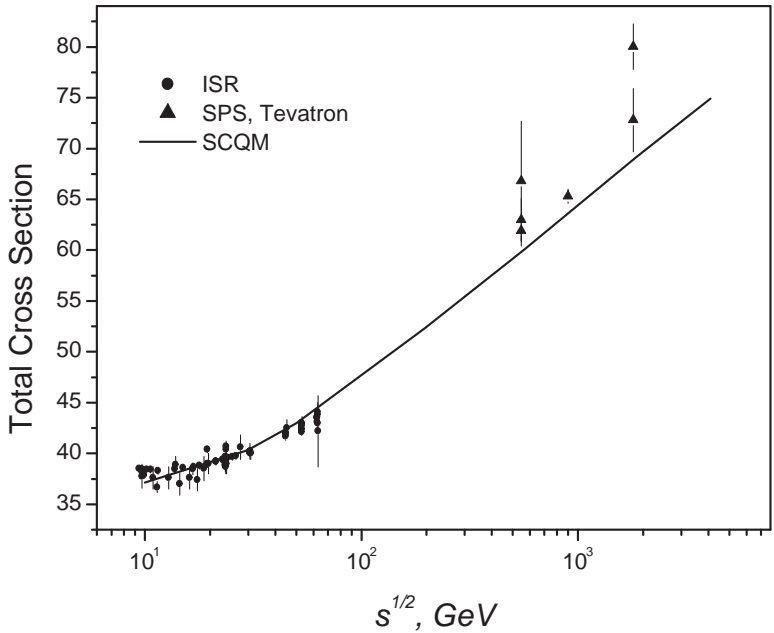}
\end{center}
\caption{Total cross section for $pp$ and $\overline{p}p-$ collisions; data
points are compilations of experimental data taken from electronic data base
HEPDATA\cite{hep}.}%
\end{figure}Since IOF relates to the probability of inelastic interaction at
given impact parameter, (12), we carried out Monte Carlo simulation of
inelastic nucleon--nucleon interactions. Inelastic interaction takes place at
definite impact parameter $b$ if at least one pion is produced in the region
where hadronic matter distributions of colliding protons overlap
\begin{equation}
4M_{q_{i}}\gamma_{q_{i}}M_{p_{j}}\gamma_{p_{j}}\int\rho_{q_{i}}(\mathbf{r}%
)\rho_{p_{j}}(\mathbf{r}-\mathbf{r}^{\prime})~d^{3}\mathbf{r}\geq
m_{\pi_{\perp}}^{2},
\end{equation}
where indices $q_{i}$ and $p_{j}$ refer to quarks from different
nucleons and $i,j=1,2,3$, $M_{q_{i}}$, $M_{p_{j}}$ -- masses of
hadronic matter composed in constituent quarks $q_{i}$ and
$p_{j}$, $\gamma_{q_{i}}$, $\gamma_{p_{j}}$ -- their
$\gamma$-factors; intergrand expression is convolution of hadronic
matter density distributions of quarks $q_{i}$ and $p_{j}.$ This
condition corresponds Hisenberg picture\cite{heisen} with modified
right hand side: $m_{\pi}^{2}$ in the original Heisenberg
inequality is replaced by $m_{\pi_{\perp}}^{2}.$ It is justified
by the fact that the average transverse momentum of produced
particles increases with energy. Specifying the quark
configurations in each colliding nucleons by (9) and (3)--(5) we
calculate $G_{in}(s,b)$ for certain values of impact parameter $b$
and then cross sections $\sigma_{in},\sigma_{el}$ and
$\sigma_{tot}.$ The values of adjusted parameters of the model,
$x_{\max}=0.64$ $fm,$ $\sigma_{x,y}=0.24$ $fm$ and
$\sigma_{z}=0.12$ $fm,$ are chosen by comparison of calculated IOF
with so called ``BEL''--parametrization\cite{henzi}, (Fig. 4), and
calculated inelastic cross sections with experimental ones,
$\sigma_{in}(s),$ in $pp$ and $\overline{p}p-$ collisions at
$\sqrt{s}_{pp}=540$ $GeV.$ Fig. 5 shows the result of calculation
for $pp$ total cross section at wide range of collision energies.
One can see that the model with parameters fixed at one energy
($540$ $GeV)$ describes the energetic behavior of $\sigma_{tot}.$
The growth of cross sections with energy is due to the continuous
tails of polarization fields around VQs not compensated by
destructive interference with fields of two other VQs. With rising
collision energy these tails result in the increasing effective
size of hadronic matter distribution inside nucleons and
correspondingly the increasing radius of interactions. The model
gives for cross sections linear logarithmic energy dependence and
the curve deviates from data points at very high energies. The
reason of this is that in our geometrical approach, ((12) --
(16)), we didn't take into account the real part of scattering
amplitude. At energies $\sqrt{s}<30$ $GeV$ calculated cross
sections were corrected on contributions of Regge poles exchange
by using
Donnachie and Landshoff parametrization\cite{donlans}:$\Delta\sigma_{R}%
^{tot}=56.08s^{-0.4526}$
\begin{figure}[tbtb]
\begin{center}
\includegraphics[width=4.0in]{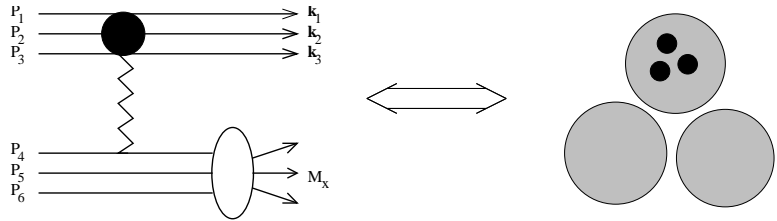}
\end{center}
\caption{Correspondence between diagram and geometrical
representation of single diffractive dissociation in
$pp\rightarrow pX$ processes. Extended sizes of colliding nucleons
are drawn on impact parameter plane: big grey circles are
constituent quarks of dissociating nucleon, small black circles --
(semi)bare
quarks of the another nucleon.}%
\end{figure}

An oscillatory motion of VQs appearing as interplay between
constituent and bare (current) quark configurations results in
fluctuations of hadronic matter distribution inside colliding
nucleons. The manifestation of these fluctuations is a variety of
scattering processes, hard and soft, in particular, the process of
single diffraction (SD). We select SD--events among inelastic
$pp\rightarrow pX$ events with the criterion $1-x_{F}<0.1$, where
$x_{F}=\frac{2}{\sqrt{s}}(k_{1}+k_{2}+k_{3})$ (Fig. 6). Here
$k_{1},$ $k_{2},$ $k_{3}$ are momenta of quarks forming the final
state proton. As one can see, SD--events correspond to constituent
quark configuration inside one colliding nucleon and (semi)bare
quark configuration inside another one. Fig. 7 shows that
calculated SD cross section slightly depends on energy; our
calculation for LHC energy gives
$\sigma_{SD}^{LHC}(pp)=8.30\pm0.15$ $mb.$
\begin{figure}[tb]
\begin{center}
\includegraphics[width=4.7in]{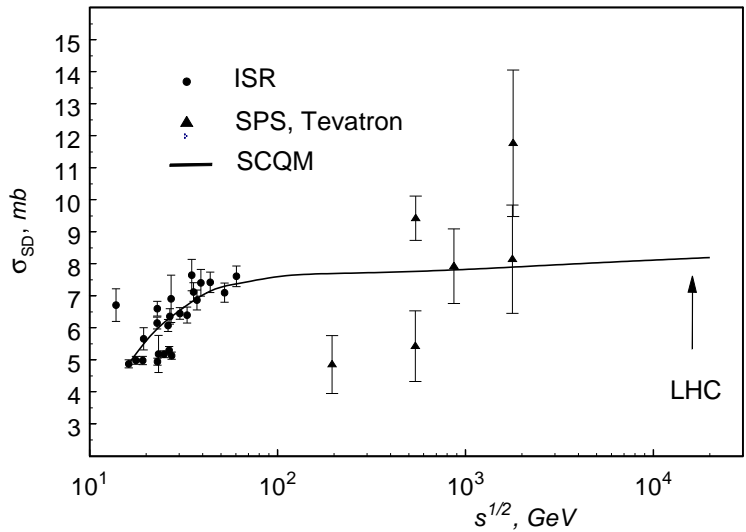}
\end{center}
\caption{Single diffractive dissociation cross section for $pp$ and
$\overline{p}p-$ collisions; data points are compilations of experimental data
taken from electronic data base HEPDATA\cite{hep}.}%
\end{figure}

\section{Discussions and Summary}

Proposed dynamical model of hadron structure, SCQM, possesses some
important features. Parameters of the model, maximal displacement
of VQ, $x_{max}=0.64 fm$, and extension of quark and gluon
condensate around VQs, $\sigma_{x,y}=0.24 fm$ and $\sigma_{z}=0.12
fm$ characterize extended sizes of hadron. Owing to noncompensated
tails of vacuum condensates at outer sides of VQs and condition
(18), the model describes the energy dependence of inelastic and
total cross sections as a result of increasing effective sizes of
colliding hadrons. On the other hand, there is no overlap of
hadronic matter distributions in space between VQs of each
interacting hadron even at very high energy collisions because
destructive interference of their polarization fields reduces
resulting hadronic matter at the center of quark system to zero
value. So, the model keeps unitarity and does not need of
inclusion of such a questionable effect as ''antishadowing''
\cite{ashad}.  According to our model diffraction dissociation is
not shadowing effect of nondiffractive inelastic process -- both
compose inelastic process, they merely differ by configurations of
quarks inside interacting hadrons.

 Because of plane oscillations of VQs and flattened (perpendicular to the plane of
oscillation) form of hadronic matter distributions nucleons are
deformed, non-spherical objects.  In paper \cite{spin}, assuming
that spin of VQs are perpendicular to the plane of oscillations,
we showed that such a deformation of proton could manifest itself
in total cross section differences between longitudinal and
transversal polarization states in proton--proton collisions.
Apparently experimental evidence of such a deformation are
quadrupole transition amplitudes in $N\rightarrow\Delta$
electroproduction which are sensitive to quadrupole deformation of
the nucleon\cite{feassler}.

Furthermore hadronic matter distribution inside hadrons and in
turn the sizes of hadrons are fluctuating quantities. Fluctuations
of extended sizes of nucleons allows one to understand why 'black
disk'' limit is not saturated in $pp$ and $\overline{p}p-$
collisions up to very high energies.

We must mention the paper\cite{barshay} of S. Barshay with
co-authors where they incorporated fluctuations in the eikonal
into geometric picture for explaining of diffractive processes.
There is a connection between eikonal and Inelastic Overlap
Function used in our calculations of cross sections through the
expression
\begin{equation}
  G_{in}(b,s)=1-e^{-2Im\chi(b,s)},
\end{equation}
were $\chi(b,s)$ is eikonal. It is obvious that fluctuations of
hadronic matter distributions caused by interplay between
constituent and current quark configurations inside colliding
hadrons lead to fluctuation of eikonal. In their approach the
fluctuations of eikonal were controlled by phenomenological
parameter, depending on energy. They predicted that single
diffraction cross section increases reaching 14--15 mb at cms
energies 20$\div$40 TeV then starts to decrease. Contrary to this
prediction there is no apparent reason for any maximum, according
to our approach, and this cross section increases steadily, very
slightly up to asymptotic energies.

 Our unified geometrical explanation of diffractive and
nondiffractive processes could give an answer on long standing
question: what is pomeron? Historically the concept of
``pomeron'', starting from simple Regge pole with intercept
$a_0=1$, transformed to a rather complicated object with
relatively arbitrary features and smooth meaning. To produce
rising cross sections it must have intercept such that
$a_0=1+\varepsilon$. The fact that the parameter $\varepsilon$ is
universal, independent of particles being scattered in hadronic
and DIS interactions, could say us that the nature of the cross
section growth is the same for all processes. Our interpretation
of pomeron is geometrical one. Both diffractive and nondiffractive
particle production emerge from disturbance (excitation) of
overlapped continuous vacuum polarization fields (gluon and
$\overline{q}q$ condensate) around valence quarks of colliding
hadrons followed by fragmentation process. The type of interaction
depends on quark configurations inside both colliding hadrons
occurring at the instant of interaction and the value of impact
parameter. So, what we used to call ``Pomeron'' in $t-$ channel is
solely continuum states in $s-$ channel and we claim that Pomeron
is unique in elastic, inelastic (diffractive and nondiffractive)
and DIS.

This research was partly supported by the Russian Foundation of Basics
Research, grants 99-07-90383 and 99-01-01103.

%

\end{document}